 \documentclass[pra,aps,twocolumn,showocis]{revtex4}
\usepackage{graphicx}
\usepackage{epsf}
\usepackage{amsmath}
\usepackage{amssymb}
\usepackage{array,multirow}
\usepackage{color}
\usepackage[cp1250]{inputenc}

\def \etal{{\em et al.}}
\def \ditto{{\em ditto}}

\begin{document}

\title{Linear-optical implementations of the iSWAP and controlled NOT
gates based on conventional detectors}

\author{Monika Bartkowiak and Adam Miranowicz}
\date{\today}

\affiliation{Faculty of Physics, Adam
 Mickiewicz University, PL-61-614 Pozna\'n, Poland}

\begin{abstract}
The majority of linear-optical nondestructive implementations of
universal quantum gates are based on single-photon resolving
detectors. We propose two implementations, which are
nondestructive (i.e., destroying only ancilla states) and work
with conventional detectors (i.e., those which do not resolve
number of photons). Moreover, we analyze a recently proposed
scheme of Wang {\em et al.}~[J. Opt. Soc. Am. B \textbf{27}, 27
(2010)] of an optical iSWAP gate based on two ancillae in Bell's
states, classical feedforward, and conventional detectors with the
total probability of success equal to $\eta^4/32$, where $\eta$ is
detector's efficiency. By observing that the iSWAP gate can be
replaced by the controlled NOT (CNOT) gate with additional
deterministic gates, we list various possible linear-optical
implementations of the iSWAP gate: (i)~assuming various ancilla
states (unentangled, two-photon and multiphoton-entangled states)
or no ancillae at all, (ii) with or without classical feedforward,
(iii) destructive or nondestructive schemes, and (iv) using
conventional or single-photon detectors. In particular, we show
how the nondestructive iSWAP gate can be implemented with the
success probability of $\eta^4/8$ assuming the same ancillae,
classical feedforward, and fewer number of conventional detectors
than those in the scheme of Wang {\em et al.} We discuss other
schemes of the nondestructive universal gates using conventional
detectors and entangled ancillae in a cluster state,
Greenberger–Horne–Zeilinger and Bell's states giving the success
probability of $\eta^4/4$, $\eta^6/8$, and $\eta^4/8$,
respectively. In the latter scheme, we analyze how detector
imperfections (dark counts in addition to finite efficiency and no
photon-number resolution) and imperfect sources of ancilla states
deteriorate the quantum gate operation.

\vspace{3mm}\noindent OCIS numbers: 270.0270, 270.5585

%\ocis{270.0270, 270.5585}

\end{abstract}

\maketitle \pagenumbering{arabic}

%------------------------------------------------------------------
\section{Introduction}

In the last decade there has been much interest in probabilistic
quantum computing using linear-optical elements and postselection
based on counts at photodetectors (see a review~\cite{Kok} and
references therein). These studies have been triggered by the
pioneering works of Knill, Laflamme, and Milburn
(KLM)~\cite{Knill} and Koashi, Yamamoto, and Imoto
(KYI)~\cite{Koashi}. Various linear-optical implementations of
universal two-qubit gates were proposed including the controlled
NOT (CNOT) and controlled sign (CS) gates as listed in Table~I.

Analysis of Table~I shows that the majority of implementations of
the CS/CNOT gates are based on selective (i.e., single-photon or
photon-number resolving) detectors and thus achieving a higher
probability of success in comparison to those schemes based on
conventional detectors. However, in practical applications the
most interesting implementations are those using conventional
detectors (also referred to as the bucket detectors) which
indicate the presence or absence of photons only.

Surprisingly, there are a very few schemes which are
nondestructive and work with conventional detectors (see Table~I).
Apart from the proposal of Zou \etal~\cite{Zou}, there are schemes
by Gasparoni \etal~\cite{Gasparoni}~(scheme \#14) and Zhao
\etal~\cite{Zhao}~(scheme \#15), which are experimental
realizations of the modified Pittman \etal~gate~\cite{Pittman}
(scheme \#12) without feedforward. In these implementations a
quantum encoder (described in Sect.~IV) was used so that the whole
setups could realize the nondestructive CNOT gate (with
single-photon detectors). However, without having such
photon-number resolving detectors for appropriate wavelength, they
used conventional detectors in experiments. Moreover, two
additional (conventional) detectors were added for postselection
of the output states. So, they only realized a {\em destructive}
version of the nondestructive CNOT gate of Pittman
\etal~\cite{Pittman}.

In Sects. III and IV, we propose two implementations of the {\em
nondestructive} universal gates based on conventional detectors.

In a recent article, Wang \etal~\cite{Wang} described a
polarization-encoded linear-optical implementation of a
nondestructive iSWAP gate using two entangled ancillae in the
Einstein-Podolsky-Rosen (EPR) states, classical feedforward and
conventional detectors. The total probability of success of this
gate is $P=\eta^4/32$, where $\eta$ is the detector efficiency and
the power of $\eta$ corresponds to the number of simultaneously
clicking detectors. In this article, we show how to simplify and
improve the scheme of Wang \etal~\cite{Wang} to obtain the
probability of success four times higher and to reduce the number
of conventional detectors, while assuming the same ancillae.

The iSWAP, CNOT and CS are universal gates, so they are formally
equivalent and each of them (together with single-qubit
operations) can be used to construct any other gates and quantum
circuits. Finding advantages of one universal gate over another
can be understood only in terms of their experimental feasibility
or specific qubit interactions in studied systems. For example, it
is usually much easier to implement the iSWAP gates rather than
the CNOT gates in solid-state systems. This is because the iSWAP
operation naturally occurs during common solid-state qubit
interactions described by the Heisenberg or XY models, while the
CNOT operation can be generated from less common Ising
interactions. For this reason, efficient quantum-information
processing based on the iSWAP gates were studied for solid-state
qubits~\cite{Tanamoto}. However, it seems that there is no clear
advantage of the linear-optical implementations of the iSWAP gates
over other universal optical gates, maybe except some realizations
in specific hybrid optical and solid-state systems.

In Sect.~II, we present simple schemes to decompose the iSWAP gate
into the CS or CNOT gate, for which many proposals (see Table~I
and Appendix~A) can be readily applied. In particular, by using
such schemes together with an implementation of the CS gate by Zou
\etal~\cite{Zou}, which was actually used in Ref.~\cite{Wang}, one
obtains the iSWAP gate with the success probability $P=\eta^4/8$.
In Sect.~III, we discuss other implementations of the iSWAP gate
yielding $P=\eta^4/4$ and $P=\eta^6/8$ using as a resource the
Gottesman-Chuang four-qubit entangled state~\cite{Gottesman} and a
pair of Greenberger–Horne–Zeilinger (GHZ) states, respectively. In
Sect.~IV, we propose a scheme using the same resources (including
ancillae in the EPR states) as the CS gate of Zou
\etal~\cite{Zou}. We conclude in Sect.~V.

%------------------------------------------------------------------
%
\begin{table*}[!ht]
\caption{List of selected linear-optical implementations of the
CS/CNOT gates, which can directly be applied to implement the
iSWAP gate. Key: $P$---the total probability of success, E
(T)---experimental (theoretical) implementation, $\vert \chi
\rangle$---the Gottesman-Chuang state equivalent to a four-qubit
cluster state~\cite{Gottesman}, $^{a}$---measurement of both the
control and target bits used for postselection, $^{b}$---assuming
perfect efficiency ($\eta=1$) of detectors. See Appendix A for
more explanations.} \vspace{.5cm}
\begin{tabular}{r c c c c c c c c}
\hline \hline\#  & Authors  & E/T  & Comments  & $P$  &
Feedforward  &
Entangled & Destructive  & {Conventional}\\
 &  & & & & & ancillae  &  & {detectors }\\
\hline
\multicolumn{9}{c}{\phantom{\Large X} I. UNENTANGLED ANCILLAE}\\
\hline 1  & KLM~\cite{Knill}  & T  &  & $\frac{1}{16}$  & no & $0$
& no  & no
\\[2pt]
2  & Ralph \etal~\cite{Ralph}  & T  & simplified \#1  & $\frac{1}{16}$  & no & 0 & no & no\\[2pt]
3  & Knill~\cite{Knill2}  & T  & improved \#1  & $\frac{2}{27}$  & no & 0 & no & no\\[2pt]
\hline
4  & Pittman \etal~\cite{Pittman4}  & E  &  & $\frac{1}{8}$  & no  & $0$  & yes$^{a}$  & no \\[2pt]
5  & \ditto  & T  & modified \#4  & $\frac{1}{4}$  & yes  & 0 & yes$^{a}$  & no\\[2pt]
\hline
6  & Giorgi \etal~\cite{Giorgi}  & T  & modified \#16  & $\frac{1}{8}$  & yes  & $0$  & no  & no\\[2pt]
\hline
7  & Bao \etal~\cite{Bao}  & E  & modified \#13  & $\frac{1}{8}$  & yes  & $0$  & no  & no \\[2pt]
\hline
\multicolumn{9}{c}{\phantom{\Large X} II. ENTANGLED ANCILLAE}\\[2pt]
\hline
8  & KLM~\cite{Knill}  & T  &  & $\frac{1}{4}$  & yes  & EPR  & no  & no\\[2pt]
\hline
9  & KYI~\cite{Koashi}  & T  &  & $\frac{1}{16}$  & yes  & EPR  & no  & no\\[2pt]
10  & \ditto  & T  & modified \#9  & $\frac{1}{4}^{b}$  & yes & $3\times$EPR  & no & no \\[2pt]
11  & \ditto  & T  & modified \#9  & $\frac{1}{4}$  & yes & $5\times$EPR  & no & no \\[2pt]
\hline
12  & Pittman \etal~\cite{Pittman}  & T  &  & $\frac{1}{16}$  & no  & EPR  & no  & no \\[2pt]
13  & \ditto  & T  & modified \#12  & $\frac{1}{4}$  & yes  & EPR  & no  & no\\[2pt]
14  & Gasparoni \etal~\cite{Gasparoni}  & E  & realization of \#12  & $\frac{1}{16}$  & no  & EPR  & yes$^{a}$  & yes\\[2pt]
15  & Zhao \etal~\cite{Zhao}  & E  & realization of \#12  & $\frac{1}{16}$  & no & EPR  & yes$^{a}$  & yes\\[2pt]
16  & Giorgi \etal~\cite{Giorgi}  & T  & related to \#12  & $\frac{1}{4}$  & yes  & EPR  & no  & no\\[2pt]
17  & Zou \etal~\cite{Zou}  & T  & related to \#12  & $\frac{1}{8}$  & yes  & $2\times$EPR  & no  & yes\\[2pt]
\hline
18 & Gottesman, Chuang~\cite{Gottesman}  & T  &   & ---  & yes  & $|\chi\rangle$  & no  & ---\\[2pt]

19  & Pittman \etal~\cite{Pittman}  & T  & based on  \#18 & $\frac{1}{4}$  & yes  & $|\chi\rangle$  & no  & no \\[2pt]
\hline
\multicolumn{9}{c}{\phantom{\Large X} III. WITHOUT ANCILLAE}\\[2pt]
\hline
20  & Pittman \etal~\cite{Pittman}  & T  &  & $\frac{1}{4}$  & no  & 0  & yes  & no \\[2pt]
21  & \ditto  & T  & modified \#20  & $\frac{1}{2}$  & yes  & 0 & yes  & no\\[2pt]
22  & Pittman \etal~\cite{Pittman2}  & E  & realization of \#20  & $\frac{1}{4}$  & no  & 0 & yes  & no\\[2pt]
23  & Giorgi \etal~\cite{Giorgi}  & T  & related to \#20  & $\frac{1}{4}$  & no  & $0$  & yes  & no \\[2pt]
24  & \ditto  & T  & modified \#23  & $\frac{1}{2}$  & yes  & 0 & yes  & no\\[2pt]
\hline
25  & Hofmann, Takeuchi~\cite{Hofmann}  & T  &  & $\frac{1}{9}$  & no  & $0$  & yes$^{a}$  & no\\[2pt]
26  & Ralph \etal~\cite{Ralph2}  & T  & equivalent to \#25  & $\frac{1}{9}$ & no  & 0 & yes$^{a}$  & no\\[2pt]
27  & O'Brien~\cite{Obrien}  & E  & realization of \#25, \#26  & $\frac{1}{9}$ & no  & 0 & yes$^{a}$  & no\\[2pt]
28  & Okamoto \etal~\cite{Okamoto}  & E  & realization of \#25, \#26 & $\frac{1}{9}$ & no  & 0 & yes$^{a}$  & no\\[2pt]
29  & Kiesel \etal~\cite{Kiesel}  & E  & simplified \#25, \#26  & $\frac{1}{9}$  & no  & $0$  & yes$^{a}$  & no \\[2pt]
30  & Langford \etal~\cite{Langford}  & E  & equivalent to \#29  & $\frac{1}{9}$ & no  & 0 & yes$^{a}$  & no \\[2pt]
 \hline\hline
\end{tabular}
\label{tabela}
\end{table*}

%------------------------------------------------------------------
\section{Decomposition of the ${\rm i}$SWAP gate and improved scheme
of Wang \etal}

%------------------------------------------------------------------
\begin{figure}[t]
\centerline{\includegraphics[width=7cm]{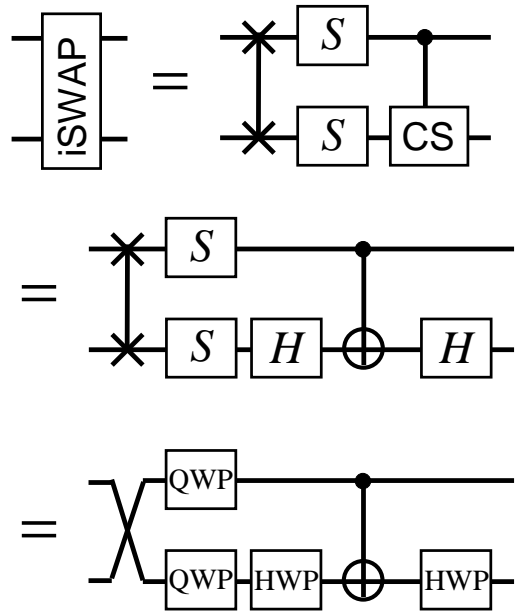}}

\caption{Circuits decomposing the iSWAP gate into the CS and CNOT
gates together with the SWAP, Hadamard ($H$) and phase ($S$)
gates. The bottom scheme shows a linear-optical realization of the
iSWAP using polarization-encoded qubits, where (i) the phase gate
is implemented (up to a global phase factor) by a quarter-wave
plate, (ii) the Hadamard gate is realized by a half-wave plate at
angle $\theta=\pi/8$, (iii) the SWAP gate can be obtained
deterministically by exchanging the qubit lines, and (iv) the
CS/CNOT can be realized probabilistically using one of the schemes
discussed in Sects.~III and~IV.}
\end{figure}

The iSWAP gate changes an arbitrary pure state of two
photon-polarization qubits
\begin{equation}
|\psi_{\rm in}\rangle=\alpha _1 \vert H H \rangle + \alpha _2
\vert H V \rangle  +\alpha _3 \vert V H \rangle  + \alpha_4  \vert
V V\rangle
 \label{eq:in}
\end{equation}
into $$|\psi_{\rm iswap}\rangle=\alpha _1 \vert H  H \rangle +
i\alpha _2 \vert V  H \rangle + i\alpha _3 \vert H V \rangle  +
\alpha_4 \vert V V\rangle,$$ where, e.g., $\vert HV \rangle=\vert
H \rangle\vert V \rangle=\vert H \rangle\otimes \vert V \rangle$
and $\vert H \rangle$ and $\vert V \rangle$ represent horizontal
and vertical polarization states, respectively. For the sake of
simplicity, we refer here to qubits encoded in photon polarization
only. Obviously, we can also refer to the photon-path and phase
qubits which are dual-line qubits interchangeable with
polarization qubits by a polarizing beam splitter and beam
splitter, respectively~\cite{Kok}.

Schuch and Siewert~\cite{Schuch} showed that the CNOT gate can be
decomposed into the two iSWAP gates or the SWAP and iSWAP gates.
The latter relation was also applied in Ref.~\cite{Wang} but not
in its full power. By inverting the Schuch-Siewert relation and
replacing the CNOT by the CS gate, we find that the iSWAP gate can
be simply given as (see the top circuit in Fig.~1)
\begin{eqnarray}
  U_{\rm iSWAP}= U_{\rm CS}(S\otimes S) U_{\rm SWAP}
\label{N1}
\end{eqnarray}
in terms of the phase gate $S={\rm diag}([1, i])$, the CS gate
$U_{\rm CS}={\rm diag}([1, 1, 1, -1])$, and the SWAP gate. The
scheme can also be given in terms of the CNOT gate, as shown in
Fig.~1 (center), using the relation $U_{\rm CS} = (I\otimes H)
U_{\rm CNOT} (I\otimes H)$. The Hadamard gate $H$ can be
implemented by the half-wave plate (HWP), which for a single qubit
is given by:
\begin{eqnarray}
  U_{\rm HWP}(\theta)=\begin{pmatrix} \cos2\theta & \sin2\theta \\
\sin2\theta & -\cos2\theta \end{pmatrix} \label{HWP}
\end{eqnarray}
tilted at $\theta=\pi/8$.

The SWAP gate is a classical gate and can be implemented
deterministically, e.g., by brute-force exchanging qubits or
waveguides carrying single qubits. Using, the polarization-encoded
qubits, the phase gate $S$ is simply implemented by a quarter-wave
plate (QWP) with fast axis horizontal.

Note that, contrary to the iSWAP gate, the entangling power of the
SWAP gate is zero, which means that this gate cannot entangle
qubits, but it is just able to alter the configuration of existing
entanglement among qubits. Sometimes this fact is confusing
because the SWAP gate is said to have a capability of two ebits,
where ebit is unit of bipartite entanglement. This is also correct
in a communication scenario.

All gates except the CS (or CNOT) are deterministic, so the
maximum success probability of the iSWAP is the same as for the CS
and CNOT.

The scheme of the iSWAP gate due to Wang \etal~\cite{Wang} is
based on proposals by Pittman \etal~\cite{Pittman} (scheme \#12 in
Table~I) and Zou \etal~\cite{Zou} (scheme \#17) implementing the
CNOT/CS gates. Scheme \#17 realizing the nondestructive CS gate
offers (to our knowledge) the highest probability of success
(equal to $1/8$) in this group of implementations using EPR states
and conventional detectors as a resource.

Thus, by applying scheme \#17 together with the decomposition
scheme shown in Fig.~1, one obtains an implementation of the iSWAP
gate yielding the probability of success $P=\eta^4/8$, which is
four times higher than that for the scheme of Wang
\etal~\cite{Wang}. Moreover, the discussed scheme requires only
eight conventional detectors instead of ten detectors used in
Ref.~\cite{Wang}.

In the next sections, we present other CNOT and CS schemes, which
can be used to implement the iSWAP gate with probability of
success equal to $\eta^4/4$, $\eta^6/8$ and $\eta^4/8$.

%------------------------------------------------------------------
\section{Scheme~I with conventional detectors and
ancillae in GHZ states}

%------------------------------------------------------------------
\begin{figure*}[!ht]
\begin{center}
\centerline{\includegraphics[width=0.8\linewidth]{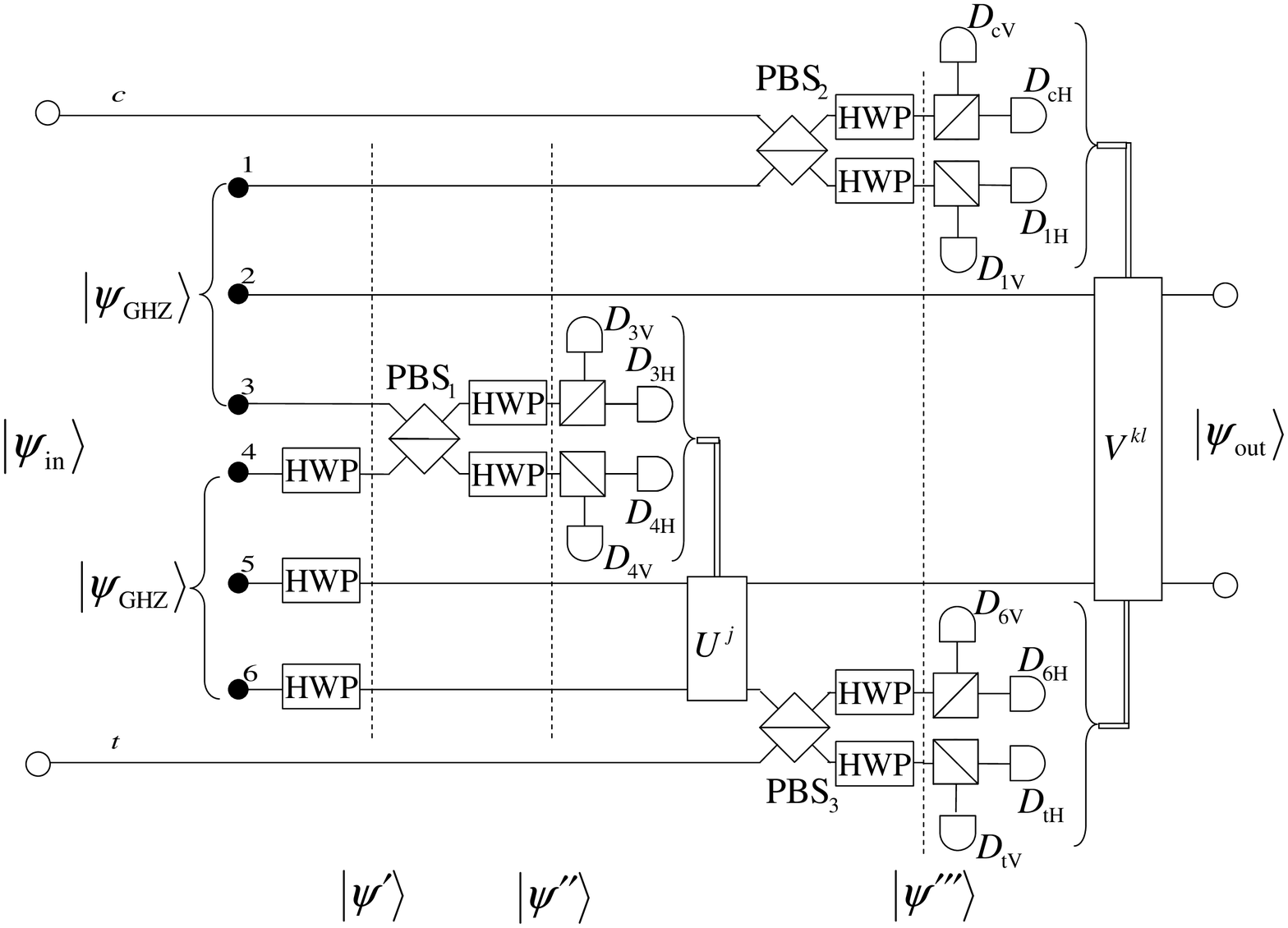}}

\caption{Scheme~I implementing the CNOT gate using conventional
detectors and ancillae in the GHZ states, $|\psi_{\rm
GHZ}\rangle$. Key: ${\rm HWP}= U_{\rm HWP}(\pi/8)$ implements the
Hadamard gate $H$; $U^{j}$ and $V^{kl}$ are conditional unitary
operations given in Table~II, where $\sigma_z$ is implemented by
$U_{\rm HWP}(0)$; $D_k$ are photodetectors; PBS$_i$ are polarizing
beam-splitters in the {\em HV}-basis.} \label{scheme1}
\end{center}
\end{figure*}
%------------------------------------------------------------------
\begin{table}[!ht]
\caption{Numbers of photons measured by ideal detectors $D_{i}$
and the corresponding required conditional operations $U^{j}$ and
$V^{kl}$ for Scheme~I.}
\begin{tabular}{c c c c c}
\hline\hline $D_{3H}$  &  $D_{3V}$ & $D_{4H}$  & $D_{4V}$ &$U^{j}$ \\
 \hline
1&0&1&0&$I$\\
0&1&0&1&$I$\\
1&0&0&1&$\sigma_z^{(5)} \otimes \sigma_z^{(6)}$ \\
0&1&1&0&$\sigma_z^{(5)} \otimes \sigma_z^{(6)}$ \\
 \hline\\
\end{tabular}
\begin{tabular}{c c c c c c c c c}
\hline $D_{cH}$  &  $D_{cV}$ & $D_{1H}$  & $D_{1V}$ &$D_{6H}$  &  $D_{6V}$ & $D_{tH}$  & $D_{tV}$ &$V^{kl}$ \\
 \hline
1&0&1&0&1&0&1&0&$I$\\
1&0&1&0&0&1&0&1&$I$\\
0&1&0&1&1&0&1&0&$I$\\
0&1&0&1&0&1&0&1&$I$\\[3pt]
1&0&0&1&1&0&1&0&$\sigma_z^{(2)} $ \\
1&0&0&1&0&1&0&1&$\sigma_z^{(2)} $ \\
0&1&1&0&1&0&1&0&$\sigma_z^{(2)} $ \\
0&1&1&0&0&1&0&1&$\sigma_z^{(2)} $ \\[3pt]
1&0&0&1&1&0&0&1&$\sigma_z^{(5)} $ \\
1&0&0&1&0&1&1&0&$\sigma_z^{(5)} $\\
0&1&1&0&1&0&0&1&$\sigma_z^{(5)} $\\
0&1&1&0&0&1&1&0&$\sigma_z^{(5)} $\\[3pt]
1&0&1&0&1&0&0&1&$\sigma_z^{(2)} \otimes \sigma_z^{(5)}$ \\
1&0&1&0&0&1&1&0&$\sigma_z^{(2)} \otimes \sigma_z^{(5)}$ \\
0&1&0&1&1&0&0&1&$\sigma_z^{(2)} \otimes \sigma_z^{(5)}$ \\
0&1&0&1&0&1&1&0&$\sigma_z^{(2)} \otimes \sigma_z^{(5)}$ \\
 \hline\hline
\end{tabular}
\end{table}

Here we describe an implementation (referred to as Scheme~I) of
the CNOT gate based on conventional detectors and ancillae
prepared in the GHZ states as shown in Fig.~2. Scheme~I is
obtained by combining the schemes of Gottesman and
Chuang~\cite{Gottesman} (scheme \#18 in Table~I) and Pittman
\etal~\cite{Pittman} (scheme \#19). It  is worth stressing that
scheme \#19 was originally designed solely for selective
detectors. Here, we show feasibility of the modified scheme \#19
using conventional detectors. Moreover, the described scheme can
be used as an implementation of the iSWAP gate according to
Fig.~1.

Schemes \#18 and \#19 use ancilla in the following cluster-type
state
\begin{eqnarray}
\vert \chi \rangle  &=& \frac{1}{\sqrt{2}}(\vert H H \rangle \vert
\Phi ^{+} \rangle    + \vert V V \rangle \vert \Psi ^{+} \rangle
), \label{chi}
\end{eqnarray}
which is equivalent (under local unitary transformations) to the
standard four-qubit cluster states~\cite{Raussendorf}. In
Eq.~(\ref{chi}), $\vert \Phi ^{+} \rangle
=\frac{1}{\sqrt{2}}(\vert H  H \rangle + \vert V  V \rangle)$ and
$\vert \Psi ^{+} \rangle =\frac{1}{\sqrt{2}}(\vert H V \rangle +
\vert V  H \rangle)$ are Bell's states (EPR states). Various
schemes for generation of the state $\vert \chi \rangle$ were
proposed including a nondestructive scheme~\cite{Wang2} yielding
the probability of success equal to $\eta^3/8$. It is possible to
generate $\vert \chi \rangle$ with the success probability
$\eta^2/2$ using the Gottesman-Chuang protocol~\cite{Gottesman},
which we apply in the following.

Our detailed implementation of the CNOT gate, as shown in Fig.~2,
is based on the schemes \#18 and \#19 and includes a scheme for
generation of the state $\vert \chi \rangle$. An arbitrary input
state $|\psi_{\rm in}\rangle$, given by Eq.~(\ref{eq:in}), is
applied in modes $c$ (control) and $t$ (target). We use two
ancillae in the GHZ states, $\vert \psi_{\rm GHZ} \rangle=
\frac{1}{\sqrt{2}}(\vert H H  H \rangle + \vert V V V \rangle)$,
as a resource. Photons in modes $4$, $5$, and $6$ are sent through
the Hadamard gate, which can be implemented by the HWP tilted at
$\theta=\pi/8$ and is described by transformations $\vert H\rangle
\rightarrow \frac{1}{\sqrt{2}}(\vert H \rangle + \vert V \rangle)$
and $\vert V\rangle \rightarrow \frac{1}{\sqrt{2}}(\vert H \rangle
- \vert V \rangle)$. For two photons with different polarizations,
the Hadamard transformation reads as $\vert H V\rangle\equiv \vert
1_H 1_V\rangle \rightarrow \frac{1}{\sqrt{2}}(\vert 2_H,0_V
\rangle - \vert 0_H,2_V \rangle)$. Thus, the total input state
(including the ancilla states) after the action of the Hadamard
gates is changed into
\begin{eqnarray}
&\vert \psi' \rangle =&\frac{1}{2\sqrt{2}}(\vert H \rangle _1 \vert H \rangle _2 \vert H \rangle _3  + \vert V \rangle _1\vert V \rangle _2\vert V \rangle _3)\nonumber \\
&& \otimes\, (\vert H \rangle _4\vert H \rangle _5\vert H \rangle _6 + \vert V \rangle _4\vert V \rangle _5\vert H \rangle _6 \nonumber \\
&&~~+ \vert V \rangle _4\vert H\rangle _5\vert V \rangle _6  +
\vert H \rangle _4\vert V \rangle _5\vert V \rangle _6).
\end{eqnarray}
The state $\vert \psi' \rangle$ is sent through polarizing
beam-splitter PBS$_1$ in the $HV$-basis (i.e., which transmits
$H$-polarized states and reflects $V$-polarized states) and the
two Hadamard gates, which results in
\begin{eqnarray}
\vert \psi'' \rangle = \frac{1}{2} \left(\vert \Phi ^{+} \rangle
_{34}U^{0} + \vert \Psi ^{+} \rangle _{34} U^{1}\right)\vert \chi
\rangle _{1256}\hspace{2.5cm}
\nonumber\\
+ \frac{1}{2}\left( \vert V \rangle _ 1 \vert V \rangle _2 \vert
\xi \rangle _{3}  \vert 0 \rangle_{4} \vert \Phi ^{+} \rangle
_{56}
%\right.    \nonumber \\ &&~~~~\left.
+ \vert H \rangle _ 1 \vert H \rangle _2 \vert 0 \rangle _{3}
\vert \xi \rangle _{4}\vert \Psi ^{+} \rangle _{56}
\right),\nonumber  \label{psi2}
\end{eqnarray}
where $U^{j}=(\sigma_z^{(5)} \otimes \sigma_z^{(6)})^j$ ($j$=0,1)
are given in terms of Pauli's matrices $\sigma_z$, $\vert \xi
\rangle =\frac{1}{\sqrt{2}}(\vert2_H \rangle - \vert 2_V
\rangle)$, and $\vert 0 \rangle\equiv\vert 0_H \rangle\vert 0_V
\rangle$ denotes no photon in $H$ and $V$ modes.

Whenever two photons reach separately detectors $D_{3H}$ and
$D_{4H}$ or $D_{3V}$ and $D_{4V}$, the state $\vert \chi \rangle$
is generated at the output (see Table~II). For combinations of
single clicks at detectors $D_{3H}$ and $D_{4V}$ or $D_{3V}$ and
$D_{4H}$, the output state requires application of two Pauli's
gates $\sigma_z$ on photons in modes $5$ and $6$ to obtain the
state $\vert \chi \rangle$. The Pauli $\sigma_z$ gate can be
implemented by the HWP at $\theta=0$ according to Eq.~(\ref{HWP}).

Thus, in the discussed part of the scheme (shown in Fig.~2 up to
$U^{j}$ operations), it is possible to generate the state $\vert
\chi \rangle$ after the successful postselection measurement and
using feedforward. The probability of success of the generation of
$\vert \chi \rangle$ is equal to $\eta^2/2$. The state $\vert \chi
\rangle$ is then used as an ancilla for the CNOT gate with the
input state $|\psi_{\rm in}\rangle$, given by Eq.~(\ref{eq:in}).

The state $|\psi''\rangle$ after measuring modes 3 and 4 and
passing through PBS$_2$ and PBS$_3$ in the $HV$-basis and four
HWPs is transformed into
\begin{eqnarray*}
|\psi''' \rangle = \frac{1}{4}[  | \Phi ^{+}\rangle _{c1} ( | \Phi
^{+}\rangle _{6t} V^{00}
+ |  \Psi ^{+}\rangle _{6t}V^{11})\hspace{2cm} \nonumber \\
 +|  \Psi ^{+}\rangle _{c1} (| \Phi ^{+}\rangle _{6t}V^{10} +|
\Psi ^{+}\rangle _{6t}V^{01})]| \psi_{\rm out} \rangle_{25} +
\frac{\sqrt{3}}{2}| \psi_{\rm err} \rangle, \label{psi3}
\end{eqnarray*}
where $V^{kl}=(\sigma_z^{(2)})^k\otimes (\sigma_z^{(5)})^l$ for
$k,l=0,1$. The state $|\psi_{\rm err} \rangle$ is a superposition
of states, which corresponds to a situation when two photons enter
one pair of detectors, $D_{iH}$ or $D_{iV}$ for some $i$
($i=c,1,6,t$). On the contrary, successful events are those, when
four photons are registered separately by all these pairs of
detectors. Conventional detectors can be used because exactly four
photons (without counting output photons) are always present in
the setup. Other cases can be easily postselected without
deteriorating the probability of success even for conventional
detectors. Because of the application of Hadamard gates in front
of polarizing beam-splitters, one can identify individual cases
and use feedforward to correct the output states when it is
necessary. After that one obtains $\vert \psi_{\rm
out}\rangle_{25}=\vert \psi_{\rm cnot}\rangle$, where $ \vert
\psi_{\rm cnot} \rangle =
  \alpha _1 \vert H H \rangle
 +\alpha _2 \vert H V \rangle
 +\alpha _3 \vert V V \rangle
 +\alpha _4 \vert V H \rangle
$ as required by the CNOT operation for the input state given by
Eq.~(\ref{eq:in}).

The probability of success of the CNOT gate is equal to $\eta^4/4$
if the state $\vert \chi \rangle$ is given. While the probability
of success for the whole scheme shown in Fig.~2, including the
generation of the state $\vert \chi \rangle $, accounts for
$\eta^6/8.$

Finally, it is worth stressing that we treat the GHZ states as a
resource. These states can be obtained from, e.g., EPR-state pairs
by applying a nondestructive optical method as proposed by
Zeilinger \etal~\cite{Zeilinger}. The first experimental
generation of the GHZ state was realized by Bouwmeester
\etal~\cite{Bouwmeester}. Since then various optical schemes for
generation of the GHZ states were described (see, e.g.,
Refs.~\cite{Fan,Sagi,Schafei,Lu}) and, in principle, such methods
can be used to generate ancillae for Scheme~I.

%------------------------------------------------------------------
\section{Scheme~II with conventional detectors and
ancillae in EPR states}

%------------------------------------------------------------------
\begin{figure*}[!ht]
\begin{center}
\includegraphics[width=0.8\linewidth]{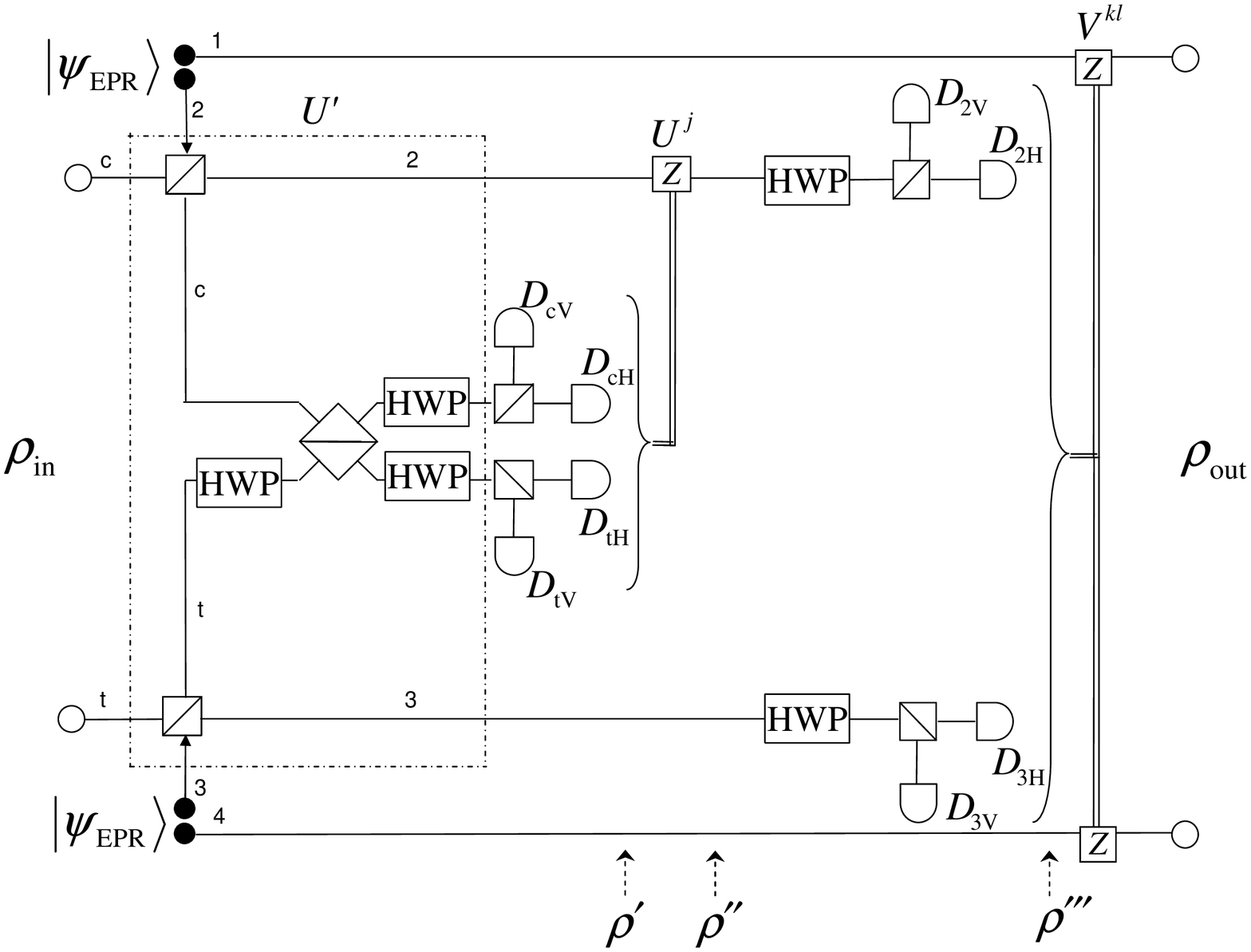}

\caption{Scheme~II implementing the CS gate using two ancillae in
perfect or non-perfect EPR states, $|\psi_{\rm EPR}\rangle$.
Notation is similar to that in Fig.~2, in particular, ${\rm HWP}=
U_{\rm HWP}(\pi/8)$ corresponds to the Hadamard gate $H$. States
and unitary operations $U'$, $U''$, $U^{j}$, and $V^{kl}$ are
defined in Sect.~IV and in Table~III.}\label{scheme2}
\end{center}
\end{figure*}
%------------------------------------------------------------------
\begin{table}[!ht]
\caption{Same as Table~II but for Scheme~II.}
\begin{tabular}{c c c c c}
\hline\hline $D_{2H}$  &  $D_{2V}$ & $D_{3H}$  & $D_{3V}$ &$U^{j}$ \\
 \hline
1&0&1&0&$I$\\
0&1&0&1&$I$\\
1&0&0&1&$\sigma_z^{(2)} $ \\
0&1&1&0&$\sigma_z^{(2)} $ \\
\hline\\
\end{tabular}

\begin{tabular}{c c c c c}
\hline $D_{cH}$  &  $D_{cV}$ & $D_{tH}$  & $D_{tV}$ &$V^{kl}$ \\
 \hline
1&0&1&0&$I$\\
0&1&1&0&$\sigma_z^{(1)} $\\
1&0&0&1&$\sigma_z^{(4)} $ \\
0&1&0&1&$\sigma_z^{(1)} \otimes \sigma_z^{(4)}$ \\
 \hline\hline
\end{tabular}
\end{table}

Here we describe an implementation of the CS gate, shown in Fig.~3
and referred to as Scheme~II, using conventional detectors and
ancillae in the EPR or EPR-like states. In our analysis of the
experimentally-oriented Scheme~II, we include a few kinds of
detector imperfections (dark counts, finite efficiency, and no
photon-number resolution) and realistic sources of the ancilla and
input states.

Our Scheme~II is a modified version of the proposals by Pittman
\etal~\cite{Pittman} (scheme \#12) and Zou \etal~\cite{Zou}
(scheme \#17). Note that scheme \#17 was also applied by Wang
\etal~\cite{Wang} as a part of their iSWAP scheme. The basic idea
of Zou \etal~was to use a quantum encoder to transform an input
state $\alpha \vert H \rangle + \beta \vert V \rangle$ into
$\alpha \vert H  H \rangle + \beta \vert V V \rangle.$ The
probability of success for such device with feedforward mechanism
is equal to $1/2$ (to compare with $1/4$ without feedforward). In
both Refs.~\cite{Zou} and~\cite{Wang} two such encoders (with
feedforward) were used to encode an input state and to obtain
finally a nondestructive gate.

Scheme~II is similar to scheme \#17 since it is also based on the
double use of the quantum encoder and the triple use of
feedforward. However, the basic idea is different: In scheme \#17,
output states of the encoders are measured separately. In
contrast, in our scheme the output states of the encoders are
combined on a PBS and only then measured. So, using this part of
Scheme~II one can generate a cluster-like state, while two
single-qubit quantum encoders in scheme \#17 can give two separate
EPR pairs. Moreover, contrary to Ref.~\cite{Zou}, we calculate
gate fidelity assuming, in particular, dark counts and realistic
sources of the EPR states.

In the case of the perfect CS gate, an arbitrary pure-state, given
by Eq.~(\ref{eq:in}), is transformed into
 $\vert \psi_{\rm cs} \rangle =
  \alpha _1 \vert H H \rangle
 +\alpha _2 \vert H V \rangle
 +\alpha _3 \vert V H \rangle
 -\alpha _4 \vert V V \rangle.
 $
Deviation of the output state $\rho_{\rm out}$ of a realistic CS
gate from the state $\vert \psi_{\rm cs} \rangle$ of an ideal CS
gate can be described by the fidelity defined by
\begin{equation}
  F=\langle \psi_{\rm cs}| \rho_{\rm out} | \psi_{\rm cs} \rangle.
\label{fidelity}
\end{equation}

Let us first analyze the action of the multigate $U'$ composed of
six gates marked in a dot-dashed box in Fig.~3:
\begin{equation}
  U' =
 U_{\rm HWP}^{(c)}
 U_{\rm HWP}^{(t)}
 U_{\rm PBS}^{(ct)}
 U_{\rm HWP}^{(t)}
 U_{\rm PBS}^{(2c)}
 U_{\rm PBS}^{(t3)},
\label{n1}
\end{equation}
where $U_{\rm PBS}^{(kl)}$ denotes the PBS unitary transformation
of $k$ and $l$ lines. The PBS operation in the dual-line
(dual-rail) notation (and assuming labelling of lines as shown
Fig.~3) corresponds to swapping of $H$-polarized modes and no
action on $V$-polarized modes. $U_{\rm HWP}=U_{\rm HWP}(\pi/8)$
corresponds to the Hadamard gate, which can be equivalently
implemented by a 50/50 beam splitter, when one of the input modes
is $H$-polarized and the other is $V$-polarized, together with two
$(-\pi/2)$ phase shifters~\cite{Kok}. The latter implementation is
particular useful to understand the Hadamard transformation
applied to more than one photon.

For a moment, let us assume that the ancillae are in the perfect
EPR states, $|\psi_{\rm EPR}\rangle=|\Phi^+\rangle$. Thus, the
total initial state is given by
$  |\Psi_{\rm in}\rangle =
  |\psi_{\rm in}\rangle_{ct}
  |\psi_{\rm EPR}\rangle_{12}
  |\psi_{\rm EPR}\rangle_{34},
 \label{n5}
$ where $|\psi_{\rm in}\rangle_{ct}$ is given by
Eq.~(\ref{eq:in}). The action of the multigate $U'$ on the initial
state $|\Psi_{\rm in}\rangle$ can be compactly written as
\begin{equation}
U'|\Psi_{\rm in} \rangle =
 N_{\rm ok} | \psi_{\rm ok} \rangle
+ N_{\rm err1} | \psi _{\rm err1} \rangle + N_{\rm err2}| \psi
_{\rm err2} \rangle , \label{new1}
\end{equation}
where
 $$|\psi_{\rm ok} \rangle=\frac{1}{\sqrt{2}}(| \Phi^{+} \rangle_{ct} U^{0}+| \Psi^{+}
\rangle_{ct} U^{1} )| \tilde C_4 \rangle_{1234}$$ with
 $U^{j}=(\sigma_z^{(2)})^j$ ($j$=0,1),
 $N_{\rm ok}^2=1/8,$
 $N^2_{\rm err1}=(8\vert\alpha_1 \vert^2+7\vert\alpha_2\vert^2+6)/16$,
 and $N^2_{\rm err2}=\vert\alpha_2
\vert^2/16+(\vert\alpha_3 \vert^2+\vert\alpha_4 \vert^2)/2$. In
general, $| \tilde C_4 \rangle_{1234}$ is of the form $
   \alpha_1 | HHHH \rangle
 + \alpha_2 | HHVV \rangle
 + \alpha_3 | VVHH \rangle
 - \alpha_4 | VVVV \rangle
\label{new3} $ which, in a special case of all equal coefficients,
reduces to a four-entangled cluster state $|C_4 \rangle$. State $
| \psi _{\rm err1} \rangle $ corresponds to undesired cases, which
{\em can} be excluded by measuring only modes $c$ and $t$ (the
first postselection). In contrast, $| \psi _{\rm err2} \rangle $
represents all the cases, in which more than one photon reaches a
detector and so, by using conventional detectors, they cannot be
distinguished from one-photon states. Thus, $ | \psi _{\rm err2}
\rangle $ corresponds to undesired cases, which {\em cannot} be
uniquely excluded via the first postselection, but {\em can} be
later excluded after measuring modes $2$ and $3$ (the second
postselection).

It is seen that, by assuming conventional detectors without dark
counts and the ancillae to be in the perfect EPR states, one
obtains the probability of success equal to $P=\eta^4/8$ and the
fidelity equal to one as in the original scheme of Zou
\etal~\cite{Zou}. Note that a successful measurement corresponds
to clicks of four out of eight detectors (see Table~III), which
explains why $P\sim\eta^4$. Moreover, factor 1/8 is just equal to
$N_{\rm ok}^2$ in Eq.~(\ref{new1}).

So far, we presented the transformations of states by assuming
perfect sources of the ancilla states and no dark counts of
detectors both for Schemes~I and~II. Here, in contrast, we use a
numerical method assuming non-perfect sources of ancillae and
input states, and dark counts.

For a conventional detector of efficiency $\eta$ and mean dark
count rate $\nu$, the positive-operator-valued measure (POVM)
elements associated with distinguishing vacuum ($\Pi _{0}$) and
the presence of at least one photon ($\Pi _ {1} $) have the form:
\begin{eqnarray}
\Pi_{0} = \sum_{m=0}^\infty e^{-\nu}(1-\eta)^{m}|m\rangle \langle
m|,\quad \Pi_{1} = 1-\Pi_{0}\, , \label{povm}
\end{eqnarray}
where $\nu=\tau_{\rm res}R_{\rm dark}$ is given in terms of the
dark count rate, $R_{\rm dark}$, and the detector resolution time,
$\tau_{\rm res}$~\cite{Ozdemir}.

We assume now that the entangled ancilla states are generated via
spontaneous parametric down-conversion (SPDC). The output state of
a type-II SPDC crystal or two type-I SPDC crystals sandwiched
together can be approximated as an EPR-like state of the form
(see, e.g., Refs.~\cite{Gerry,Gilchrist}):
\begin{equation}
  |\psi_{\rm EPR}\rangle = (1-\gamma^2)^{-1/2}
  [|0\rangle|0\rangle+\gamma(|HH\rangle+|VV\rangle)]
  +{\cal O}(\gamma^2),
 \label{n7}
\end{equation}
where parameter $\gamma$ is given by the product of interaction
time of the pump field and the crystal, their coupling constant,
and complex amplitude of the pump field. State, given by
Eq.~(\ref{n7}), clearly differs from the exact EPR state
$|\Phi^+\rangle$ by inclusion of vacuum (and also higher
order-states) in the superposition. Parameter $\gamma^{2}$ is
usually of the order $10^{-4}$/pulse~\cite{Ozdemir} and it
describes the rate of single-photon pair generation per pulse of
the pump field. Thus, the output state of the SPDC crystal
contains vacuum with high probability and its effect on the gate
operation cannot be neglected.

Each line in Schemes~I and~II can carry arbitrary number of
photons in $H$ and $V$ polarizations. Using a dual-line notation,
one can write
 $|H\rangle = |1\rangle_H|0\rangle_V\equiv |1_H, 0_V\rangle,$
 $|V\rangle = |0\rangle_H|1\rangle_V,$ and
 $|0\rangle = |0\rangle_H|0\rangle_V$.

The state $\rho'$ after the action of the multigate $U'$ and the
measurement of photons by the detectors $D_{cH}$, $D_{cV}$,
$D_{tH}$, and $D_{tV}$ is given by:
\begin{equation}
  \rho' = {\cal N}\,{\rm Tr}_{ct} \left[
  \Pi_m^{(cH)}\Pi_{m'}^{(cV)}\Pi_n^{(tH)}\Pi_{n'}^{(tV)}
  U'\rho_{\rm in} (U')^\dagger \right],
\label{n3}
\end{equation}
where ${\rm Tr}_{ct}\equiv {\rm Tr}_{cH,cV,tH,tV}$, $\rho_{\rm
in}=|\Psi_{\rm in}\rangle\langle\Psi_{\rm in}|$, ${\cal N}$ is a
renormalization constant, and the POVM elements are given by
Eq.~(\ref{povm}). Moreover, $m,m',n,$ and $n'$ are equal to 1 or
0, corresponding to clicks or no clicks of the detectors according
to Table~III.A. By applying the conditional gate
$U^j=(\sigma_z^{(2)})^j$ with $j=0,1$, defined in Table~III.A, the
state $\rho'$ is transformed to $\rho''=U^j\rho' (U^j)^ \dagger$.
After the operation $U'' = U_{\rm HWP}^{(2)} U_{\rm HWP}^{(3)}$
corresponding to the Hadamard gates at lines 2 and 3, and after
photon counting by the detectors $D_{2H}$, $D_{2V}$, $D_{3H}$, and
$D_{3V}$, the state $\rho''$ is transformed to
\begin{equation}
  \rho''' = {\cal N}\,{\rm Tr}_{23} \left[
  \Pi_{m}^{(2H)}\Pi_{m'}^{(2V)}\Pi_{n}^{(3H)}\Pi_{n'}^{(3V)}
  U''\rho'' (U'')^\dagger\right],
\label{n4}
\end{equation}
where ${\rm Tr}_{23}\equiv {\rm Tr}_{2H,2V,3H,3V}$, while
$m,n,m'$, and $n'$ correspond to clicks or no clicks of the
detectors according to Table III.B. Note that the PBSs in front of
all the detectors just convert polarization qubits into dual-line
qubits, so they are redundant if we apply the dual-line notation
consistently in our numerical approach. The final output state
$\rho_{\rm out}=V^{kl} \rho'' (V^{kl})^{\dagger}$ is obtained from
$\rho''$ by applying the conditional gates
$V^{kl}=(\sigma_z^{(1)})^k \otimes (\sigma_z^{(4)})^l$ ($k,l=0,1$)
according to Table III.B.

For simplicity, in our numerical calculations we reserved
three-dimensional Hilbert space for each mode, thus we set
$|0\rangle_H=[1;0;0]$, $|1\rangle_H=[0;1;0]$, and
$|2\rangle_H=[0;0;1],$ and analogously for $V$ polarization. This
is valid by assuming dark count rates and $\gamma$ parameter to be
relatively low. Otherwise, higher-dimensional Hilbert spaces
should be set.

Let us assume realistic values of conventional
detectors~\cite{CPC1} (see also Refs.~\cite{Ozdemir,Miran}): the
detector efficiency to be $\eta=0.7$, the dark count rate $R_{\rm
dark}=100~{\rm s}^{-1}$, the detector resolution time $\tau_{\rm
res}$=10 ns. For convenience, we assume that all detectors are the
same. The rate of single-photon pair generation per pulse of the
pump field is set to be $\gamma^{2}=10^{-4}$/pulse~\cite{Ozdemir}.
For experimental verification of Scheme~II, it is useful to assume
that the input state $|\psi_{\rm in}\rangle$ is also generated by
the SPDC and is given by Eq.~(\ref{n7}). For brevity, we analyze
only the first cases in Table~III, where no extra conditional
operations are required. Under these assumptions, we find that the
fidelity drops to $F\approx 0.97$, which is still relatively high.

%------------------------------------------------------------------
\section{Conclusions}

We studied linear-optical implementations of two-qubit universal
gates including the iSWAP and CS/CNOT gates. As shown in Table~I,
the majority of these realizations of nondestructive gates are
based on single-photon detectors. In contrast, we focused on
practical implementations using conventional detectors, which do
not resolve number of photons.

Despite of progress in constructing single-photon detectors (see
Refs.~\cite{JMO,Takeuchi} and references therein), they are still
not commonly used. This conclusion can be drawn, e.g., by
analyzing experimental realizations of quantum gates listed in
Table~I. One of the drawbacks of single-photon detectors is that
their dark count rates are much higher than those for conventional
detectors~\cite{Takeuchi}. There are also proposals of
multiple-photon resolving detectors including cascade arrays of
conventional detectors (connected with beam splitters or with
high-speed low-loss optical switches~\cite{Castelletto}) and
fiber-loop detectors~\cite{Rehacek}. Such detectors, which are
based on the idea of chopping up photons, are conceptually very
attractive but still experimentally underdeveloped.

We analyzed a recent proposal of Wang \etal~\cite{Wang} to
implement the iSWAP gate using two entangled ancillae in EPR
states, classical feedforward, and conventional photodetectors (of
a finite efficiency $\eta$) with the success probability of
$\eta^4/32$ only. This scheme was based on an implementation of
the CS gate by Zou \etal~\cite{Zou} (scheme \#17 in Table~I) with
the success probability of $\eta^4/8$.

We showed that the iSWAP gate can be decomposed into the CS/CNOT
gate and deterministic gates including the SWAP, phase or Hadamard
gates. Thus, one can immediately obtain schemes that implement the
iSWAP gate by using the CS/CNOT gates with relatively high
probability of success. In particular, by applying scheme \#17 of
Zou \etal~\cite{Zou} together with the iSWAP decomposition scheme,
we showed how to implement the iSWAP gate with the success
probability four times higher than that in the Wang \etal~scheme.

Moreover, we studied applicability of conventional detectors to
other implementations of nondestructive gates originally designed
for single-photon detectors. We showed that the scheme of Pittman
\etal~\cite{Pittman} implementing the nondestructive CNOT gate can
be used also with conventional detectors achieving the probability
of success equal to $\eta^4/4$ assuming as a resource the
Gottesman-Chuang four-qubit entangled state~\cite{Gottesman} or
equal to $\eta^6/8$ for a pair of ancillae in the GHZ states.

We have also described another scheme based on conventional
detectors and ancillae in the EPR or EPR-like states as a modified
version of the scheme by Zou \etal~\cite{Zou}. To verify
experimental feasibility of this scheme, we showed how the quantum
gate fidelity is deteriorated due to realistic sources of ancilla
and input states, and detector imperfections to include dark
counts, finite efficiency and no photon-number resolution.

%------------------------------------------------------------------

\noindent {\bf Acknowledgements.} The authors thank Nobuyuki
Imoto, Bryan Jacobs, Masato Koashi, and Zhi Zhao for discussions.
The work was supported by the Polish Ministry of Science and
Higher Education under Grant No. 2619/B/H03/2010/38.

\begin{appendix}

%------------------------------------------------------------------
\section{Comparison of the schemes listed in Table~I}

Here, we give more explanations and compare various linear-optical
implementations of the CS/CNOT gates listed in Table~I. Obviously,
these schemes can be used also to construct the iSWAP gate
according to Fig.~1.

The implementations can be divided into several groups according
to, e.g., different resources as shown in Table~I: (I) unentangled
ancillae, (II) entangled ancillae, and (III) without ancillae at
all. Our examples of the second group include ancillae in the EPR
states described in Sect.~IV, but also the Gottesman-Chuang
four-entangled state and the GHZ states discussed in Sect.~III.

We compared the schemes concerning the total probability of
success, destructive or nondestructive character of the
implementations, application of conventional or nonconventional
detectors, and whether the feedforward mechanism was applied.
Classical feedforward means that a scheme includes measurement
devices of some modes such that the classical outcomes of the
measurements can be used to change the remaining modes.

In group I, where one or two ancillae prepared in an unentangled
state were used, the highest probability of success for the gates
without feedforward accounts for $2/27$~\cite{Knill2}~(for scheme
\#3 in Table~I). It is worth noting that there is only a numerical
evidence~\cite{Uskov}, but not an analytical proof (contrary,
e.g., the nonlinear sign shift gate~\cite{Eisert}) that $2/27$ is
the rigorous tight upper bound on the success probability using
two unentangled ancillae without feedforward. Moreover, additional
ancillae do not increase this value. When feedforward is used the
probability can be increased to $1/8$ for gates with two
ancillae~\cite{Giorgi,Bao}~(schemes \#6 and \#7) or even to $1/4$
with one ancilla~\cite{Pittman4}~(schemes \#4 and \#5) at the
expense of destructing the output states. It should be mentioned
that for all these groups of implementations, the destructive
gates (i.e., those for which {\em not} only ancilla states are
measured) achieve higher probabilities.

In group II, the best  achieved probability of success accounts
for $1/16$ without feedforward~\cite{Koashi,Pittman}~(schemes \#9
and \#12) and $1/4$ with
feedforward~\cite{Knill,Pittman,Giorgi}~(schemes \#8, \#13, and
\#16).

Group III consists of the CS/CNOT gates based on the idea of
Hofmann and Takeuchi~\cite{Hofmann}~(scheme \#25), and Ralph
\etal~\cite{Ralph2}~(scheme \#26). Other examples in this group
are mainly experimental realizations of the schemes \#25 and \#26
using a beam splitter with the reflection coefficient equal to
$1/3$. The probability of success for them achieves $1/9$,
assuming the measurement of both the control and target bits for
postselection.

Intentionally, we have not included implementations of the CS/CNOT
gates based on the idea of one-way computation using cluster
states as proposed by Raussendorf and Briegel~\cite{Raussendorf}.
According to their proposal one can implement the CS/CNOT gate by
performing single-qubit measurement in an appropriate basis on a
given cluster state. Using this procedure with additional
feedforward it is possible to implement the CS/CNOT gate nearly
deterministically even with conventional detectors as described,
e.g., in Refs.~\cite{Yoran,Nielsen,Browne,Tokunaga} and
experimentally realized in
Refs.~\cite{Walther,Kiesel2,Tokunaga2,Gao}.

However, it should be stressed that such implementations of the
CS/CNOT gates based on cluster-type states look deterministic only
because it is assumed something strictly easier than applying the
true CNOT gate on {\em independently} prepared input photonic
qubits. The latter task should not be deterministic because of the
no-go theorem for the Bell measurement by linear optics.

It worth clarifying that Table~I includes two schemes using the
cluster-type states. Namely, schemes of Gottesman and
Chuang~\cite{Gottesman}~(scheme \#18) and closely related proposal
of Pittman \etal~\cite{Pittman}~(scheme \#19) are implementations
of the nondestructive and nondeterministic CNOT gate using a
four-photon entangled state $|\chi\rangle$, which is equivalent,
under a local unitary transformation, to a four-qubit cluster
state. We included this gate in Table~I since it does not realize
the Raussendorf-Briegel protocol but uses the state $\vert \chi
\rangle$ as an ancilla only.

In Table~I, we also have not included deterministic
implementations of the universal gates based on single-photon
cross-Kerr nonlinearities (see Refs.~\cite{Nemoto,Kok} and
references therein). Such schemes are fundamentally different from
probabilistic linear-optical schemes. Moreover, there are serious
doubts~\cite{Shapiro} whether they can be useful for quantum
computing if applied for single photons in a standard way.

\end{appendix}

%------------------------------------------------------------------

\end{document}